\documentclass[12pt]{article}

\usepackage{scicite}

\usepackage{times}

\usepackage{hyperref}
\usepackage{amsmath}
\usepackage{amsfonts}
\usepackage{amssymb}
\usepackage{graphicx}
\usepackage[FIGTOPCAP,raggedright,nooneline]{subfigure}
\usepackage{float}
\usepackage{longtable}
\usepackage{subfigure}
\usepackage{amsfonts}
\usepackage{amsthm}
\usepackage{bbold}
\usepackage[normalem]{ulem}
\usepackage{comment}
\usepackage{hyperref}
\usepackage{multirow}%
\usepackage{amsmath,amssymb,amsfonts}%
\usepackage{amsthm}%
\usepackage{mathrsfs}%
\usepackage{booktabs}%
\usepackage{lineno}
\usepackage{tablefootnote}
\usepackage{caption}
\usepackage[title]{appendix}%
\usepackage{xcolor}%
\usepackage{manyfoot}%
\usepackage{hyperref}
\usepackage{afterpage}
\usepackage{chapterbib}
\hypersetup{colorlinks = true, linkcolor = black}
\newenvironment{sciabstract}{%
\begin{quote} \bf}
{\end{quote}}

\newcommand{\ket}[1]{\left\vert#1\right\rangle}
\newcommand{\bra}[1]{\left\langle#1\right\vert}

\title{Experimental demonstration of generalized quantum fluctuation theorems in the presence of coherence}

\author
{Hui Li${}^{1 \dagger},$ Jie Xie${}^{1 \dagger},$ Hyukjoon Kwon${}^{2, 3 \dagger},$ Yixin Zhao${}^{4},$ M.S. Kim${}^{3 \ast},$ Lijian Zhang${}^{1 \ast}$\\
\\
\normalsize{${}^{1}$National Laboratory of Solid State Microstructures, }\\
\normalsize{Key Laboratory of Intelligent Optical Sensing and Manipulation, }\\
\normalsize{and Collaborative Innovation Center of Advanced Microstructures, }\\
\normalsize{College of Engineering and Applied Sciences and School of Physics,}\\
\normalsize{Jiangsu Physical Science Research Center,}\\
\normalsize{Nanjing University, Nanjing, 210093, China}\\
\normalsize{${}^{2}$School of Computational Sciences, Korea Institute for Advanced Study,}\\
\normalsize{Seoul, 02455, Korea}\\
\normalsize{${}^{3}$Blackett Laboratory, Imperial College London, London,}\\
\normalsize{SW7 2AZ, United Kingdom}\\
\normalsize{${}^{4}$School of Electronics, Center for Quantum Information Technology,}\\
\normalsize{Peking University, Beijing, 100871, China}
\\
\normalsize{$^\dagger$These authors contributed equally to this work.}\\
\normalsize{$^\ast$Corresponding author. E-mail: lijian.zhang@nju.edu.cn (L.J.Zhang)}\\
\normalsize{$^\ast$Corresponding author. E-mail: m.kim@imperial.ac.uk (M.S.Kim)}\\
}

\date{}

\begin{document} 
\captionsetup[table]{labelfont={bf},name={Table},labelsep=period}
\captionsetup[figure]{labelfont={bf},name={Fig.},labelsep=period}

\baselineskip24pt


\maketitle



\begin{sciabstract}
				Fluctuation theorems have elevated the second law of thermodynamics to a statistical realm by establishing a connection between time-forward and time-reversal probabilities, providing invaluable insight into non-equilibrium dynamics. While well-established in classical systems, their quantum generalization, incorporating coherence and the diversity of quantum noise, remains open. We report the experimental validation of a quantum fluctuation theorem (QFT) in a photonic system, applicable to general quantum processes with non-classical characteristics, including quasi-probabilistic descriptions of entropy production and multiple time-reversal processes. Our experiment confirms that the ratio between the quasi-probabilities of the time-forward and any multiple time-reversal processes obeys a generalized Crooks QFT. Moreover, coherence induced by a quantum process leads to the imaginary components of quantum entropy production, governing the phase factor in the QFT. These findings underscore the fundamental symmetry between a general quantum process and its time reversal, providing an elementary toolkit to explore noisy quantum information processing.
\end{sciabstract}

\section*{Teaser:}Experimental demonstration of generalized quantum fluctuation theorems incorporating the effects of coherence and noise.

\section*{Introduction}
Irreversibility is a universal and predominantly unavoidable feature of nature. In thermodynamics, this feature is explained through the concept of entropy production, described in terms of the system's entropy change and heat exchange with the surrounding environment. While the second law of thermodynamics states that entropy production is non-decreasing on average, the development of fluctuation theorems (FTs) allows for a deeper understanding of thermodynamic quantities under non-equilibrium processes beyond their average behaviors. Especially, the Crooks FT establishes a fundamental symmetry between the probability distributions of entropy production $\omega$ for the forward ($P_{\rightarrow}\left( \omega \right)$) and time-reversal ($P_{\leftarrow}\left( -\omega \right)$) processes \cite{PhysRevE.60.2721}
\begin{equation}\label{Eq:Crooks_FT}
	\frac{P_{\rightarrow}(\omega)}{P_{\leftarrow}(-\omega)}=e^{\omega},
\end{equation}
which leads to the integral FT $\overline{ e^{-\omega} } = 1$ by averaging over all possible trajectories~\cite{PhysRevLett.95.040602}. As being equality conditions containing all the high-order moments, FTs provide a precise description of non-equilibrium dynamics in the microscopic scale, which readily implies the second law of thermodynamics ($\overline \omega \geq 0$), where $\overline{\omega}$ is the average entropy production in the macroscopic scale. For the past two decades, FTs for classical systems~\cite{PhysRevLett.78.2690,PhysRevE.60.2721,RevModPhys.81.1665,RevModPhys.83.771} have achieved a great success for understanding the irreversibility in non-equilibrium dynamics of a wide range of systems from biological systems to nano-scale heat engines~\cite{Liphardt1832,collin_verification_2005,PhysRevLett.96.070603,PhysRevLett.99.068101,toyabe_experimental_2010,PhysRevLett.109.180601}.

With the development of quantum information science along with the precise control of quantum systems, there has been a demand for the quantum generalization of FT to better understand the non-equilibrium dynamics in quantum devices. However, extending FTs to the quantum regime presents considerable challenges, as work and heat are not well defined in quantum systems due to the existence of coherence~\cite{PhysRevE.75.050102}, which is the notable difference between classical and quantum mechanics. A systematic way to define thermodynamic quantities with two-point measurement (TPM), i.e., measurements performed both before and after the evolution of the system, is extended to quantum systems, from which classical FTs can be recovered in the language of quantum mechanics \cite{RevModPhys.81.1665,RevModPhys.83.771,RevModPhys.93.035008} and have been verified experimentally \cite{PhysRevLett.101.070403,an_experimental_2015,Smith_2018,PhysRevLett.113.140601,cerisola_using_2017,Zhang_2018,PRXQuantum.2.030353,PhysRevA.106.L020201,PhysRevLett.129.170604}. Despite the benefits of the TPM approach, its projective nature represents a fundamental limitation: it irreversibly destroys the coherence in initial states, thereby prohibiting investigations of the role of coherence in the QFTs~\cite{PhysRevE.75.050102,PhysRevE.90.032137,PhysRevLett.118.070601}. To address this problem, considerable efforts have been devoted to understanding the role of coherence in thermodynamics~\cite{Wueaav4944,PhysRevE.92.042150,PhysRevA.103.L060202,PhysRevA.97.052122,PhysRevA.102.042220,santos_role_2019} and establishing a fully quantum version of FTs~\cite{PhysRevE.92.042113,PhysRevLett.124.090602, Aberg2018FullyQuantum, Holmes2019coherentfluctuation}.

Another intriguing direction in both classical and quantum FTs is to generalize the theory to be applied to a wider class of non-equilibrium processes \cite{PhysRevLett.86.3463,PhysRevLett.104.090601, PhysRevX.8.031037}, which can involve coherence in quantum channels. This generalization holds particular significance in quantum information processing as noisy quantum operations can go beyond the description of the thermodynamic process. In this vein, the QFTs that can be applied to a quantum system coupling to the environment in a more general way \cite{PhysRevX.8.031037, PhysRevLett.127.180603,hernandez2023experimental,rodrigues2024nonequilibrium} were recently proposed and demonstrated. However, these methods necessitate the measurement on the environment, which may not be feasible in most situations, or require a specific condition on system-environment coupling.

In this article, we explore QFTs that can be applied to an arbitrary quantum channel, which was introduced in Ref.~\cite{PhysRevX.9.031029}. 
A distinctive feature of quantum systems is that coherence—captured by the off-diagonal elements of the density matrix—also fluctuates during a quantum process. As these off-diagonal elements are generally complex-valued, the corresponding fluctuation theorems may be formulated to incorporate complex values, extending their classical counterparts. Building on this idea, the fluctuation theorem presented in Ref.~\cite{PhysRevX.9.031029}
involves the quasi-probability $P_{\rightarrow \left( \leftarrow \right)}\left( \omega \right) $ and complex-valued entropy production $\omega = \omega_{R}+i\omega_{I}$ as
\begin{equation}\label{Eq:Crooks_FT_Complex}
				\frac{P_{\rightarrow}(\omega)}{P^\theta_{\leftarrow}(-\omega^*)}=e^{\omega_R - 2 i \theta \omega_I},
\end{equation}
where $\omega^*$ is the complex conjugate of $\omega$. Here, the probability distribution of entropy production is generalised to the quasi-probability distribution which can have non-real values to fully incorporate the effect of coherence both within states and during the quantum process.
Our construction of the complex-valued quasi-probability shares a common structure with the Kirkwood-Dirac distribution~\cite{Kirkwood33, Dirac45}, which has been widely explored~\cite{PhysRevE.90.032137, PhysRevLett.120.040602, PRXQuantum.1.010309, zhang2024quasiprobability} as a possible replacement for the TPM scheme in QFTs~\cite{PhysRevLett.118.070601}. The occurrence of negative or non-real values in quasi-probability distributions indicates the non-classical nature of quantum systems when considering the joint distribution of incompatible quantum observables~\cite{Lostaglio2023kirkwooddirac}.
In our case, quantum observables corresponding to entropy and heat
are incompatible as they do not commute, in general.
Another important characteristic is that depending on the time-translation symmetry of the quantum process~\cite{petz1986sufficient,junge_universal_2018}, its time reversal may not be uniquely defined for a quantum process. Different choices of the time-reversal process can be parameterized by a rotational degree of freedom (DOFs) $\theta$, which will be described in more details later. We note that the imaginary component $\omega_{I}$ in Eq.~\eqref{Eq:Crooks_FT_Complex} is closely related to the phase factor arising from multiple time-reversal processes.

We experimentally test the validity of the QFT in Eq.~\eqref{Eq:Crooks_FT_Complex} for a non-unitary quantum channel with a quantum photonic system. Such systems have been demonstrated as versatile tools for the investigation of quantum thermodynamics recently~\cite{PhysRevLett.116.050401,PhysRevLett.118.130502,PhysRevLett.121.160602,tham2016simulating, Araujo_2018}. The quasi-probability distributions of the quantum entropy production are reconstructed for both forward and time-reversal processes based on a two-point generalized measurement protocol. The experimental data clearly follows the QFT for both covariant (with time-translation symmetry) and incovariant (with broken time-translation symmetry) quantum channels. Compared to other QFTs focusing on covariant channels~\cite{Aberg2018FullyQuantum, PhysRevX.8.031037}, we test a more general form of QFTs, in which the imaginary entropy production plays a crucial role in capturing fully quantum effects during incovariant coherence transitions. Our experimental results demonstrate the universal relationship between a quantum channel and its time-reversal channels, setting fundamental limitations on the reversibility of quantum operations.

\section*{Results}
\subsection*{Quantum entropy production and multiple time-reversal processes}
We characterize the quantum entropy production through a general quantum channel ${\cal N}$ in terms of the initial state $\hat\rho^I = \sum_{\mu}p_{\mu}^{I}|\phi_{\mu}^{I}\rangle \langle \phi_{\mu}^{I}\vert = \sum_{\mu}p^{I}_{\mu}\hat{\Phi}_{\mu}^{I}$ and the final state $\hat\rho^F = {\cal N}(\hat\rho^I) = \sum_{\nu}p_{\nu}^{F}\vert\phi_{\nu}^{F}\rangle \langle \phi_{\nu}^{F}\vert = \sum_{\nu}p^{F}_{\nu}\hat{\Phi}^{F}_{\nu}$. We assume that $\hat{\gamma} = \sum_{i}r_{i}\vert i\rangle \langle i \vert =\sum_{i}r_{i}\hat{\Pi}_{i}$ is the channel's stationary state such that ${\cal N}(\hat\gamma) = \hat\gamma$.
By introducing the von Neumann entropy $S(\hat\rho) = -{\rm Tr}[\hat\rho \ln \hat\rho]$ and a non-equilibrium potential $(-\ln\hat\gamma)$, average entropy production can be generally defined as $\overline\omega = \Delta S - {\rm Tr} [(\hat\rho^F - \hat\rho^I) (-\ln \hat\gamma)] $~\cite{PhysRevX.8.031037}. For a thermal channel in contact with the heat bath with temperature $T$, this definition recovers the conventional entropy production, $\overline\omega = \Delta S - Q/T$ with $Q$ being heat exchange, by taking $\hat{\gamma} \propto e^{-\hat{H}/T}$, the equilibrium state when the Hamiltonian of the system is $\hat{H}$. Consequently, $\overline \omega \geq 0$ is regarded as a quantum generalization of the second law of thermodynamics.

The fluctuation of the quantum entropy production can then be explored by considering transitions between the eigenstates, $\{\hat\Phi_\mu^I\} \rightarrow  \{\hat\Phi_\nu^F\}$, with the probability $T^{\mu \rightarrow \nu} = {\rm Tr} [{\cal N} (\hat\Phi_\mu^I) \hat\Phi^F_\nu]$. However, a critical issue arises when the quantum operator corresponding to heat, or more generally, the non-equilibrium potential $(-\ln \hat\gamma)$ does not commute with $\hat\rho^I$ or $\hat\rho^F$. In this case, the transition $\hat\Phi^I_\mu \rightarrow \hat\Phi^F_\nu$ may contain the transitions between the off-diagonal elements $\ket{i}\bra{j} \rightarrow \ket{k}\bra{l}$ with respect to the eigenstates of the stationary state $\hat\gamma$, and this transition cannot be described by classical probability. Nevertheless, this problem can be detoured by introducing a complex-valued transition amplitude $T^{\mu\to \nu}_{ij\to kl}$ between operators $\hat{\Pi}_i \hat{\Phi}^I_\mu \hat\Pi_j$ and $\hat{\Pi}_k \hat{\Phi}^F_{\nu} \hat{\Pi}_l$ (see Methods). For each transition, the complex-valued entropy production can be written as~\cite{PhysRevX.9.031029}
\begin{equation}
\omega^{\mu  \rightarrow \nu}_{ij \rightarrow kl}  =  \ln \left( \frac{p^I_\mu \sqrt{r_k r_l} }{p^F_{\nu}\sqrt{r_i r_j} }\right) + i \ln \left( \frac{\sqrt{r_j r_l} }{\sqrt{r_i r_k} }\right).
\label{eq:ent_prod}
\end{equation}

Now we define the distribution of the quantum entropy production $P_\rightarrow(\omega)$, which correctly indicates the average entropy production $\sum_\omega \omega P_\rightarrow(\omega) = \overline\omega$,
where $\omega = \omega_R + i\omega_I$ and $\sum_\omega$ denotes the summation over $\omega_R$ and $\omega_I$ (see Methods). 
Such a distribution can be constructed in terms of the transition amplitude as~\cite{PhysRevX.9.031029}
\begin{equation}
P_\rightarrow(\omega) = \sum_{\mu,\nu,i,j,k,l} \delta(\omega - \omega^{\mu \rightarrow \nu}_{ij \rightarrow kl} ) p^I_\mu T^{\mu \rightarrow  \nu}_{ij \rightarrow kl}.
\label{eq:entprod_dist}
\end{equation}

A major difference in the quantum entropy production Eq.~\eqref{eq:ent_prod} compared to its classical counterpart in Eq.~\eqref{Eq:Crooks_FT} is that the quantum entropy production can be complex-valued as well as its quasi-probability distribution. However, it is not always the case that a quantum process results in complex-valued entropy production. A trivial case is when $\hat\rho^I$, $\hat\rho^F$, and $\hat\gamma$ have common eigenstates, in which case Eq.~\eqref{Eq:Crooks_FT_Complex} reduces to the classical Crooks FT in Eq.~\eqref{Eq:Crooks_FT} as such a process can be regarded as a classical stochastic process. A non-trivial case happens when a quantum channel is covariant under group transformation $\hat{U}_{\hat\gamma} (\theta) = e^{-i \theta \ln\hat\gamma}$, parameterized by a rotational DOF $\theta$, satisfying
\begin{equation} \label{eq:covariance_cond}
{\cal N}(\hat{U}_{\hat\gamma}(\theta) \hat{\rho} \hat{U}^\dagger_{\hat\gamma}(\theta) ) = \hat{U}_{\hat\gamma}(\theta) {\cal N}(\hat{\rho})\hat{U}^\dagger_{\hat\gamma}(\theta)\quad \forall \theta \in \mathbb{R},
\end{equation}
in which case, both entropy production $\omega$ and $P_\rightarrow(\omega)$ are real-valued~\cite{PhysRevX.9.031029}. For a thermodynamic process, this condition coincides with the time-translation symmetry generated by the system Hamiltonian $\hat{H}$. Conversely, an incovariant quantum channel can be witnessed by the imaginary part of entropy production (see Table. \ref{tab1}).

\begin{table}[ht]
\centering
\caption{\textbf{Quantum entropy production $\left( \omega \right) $ and time-reversal channel for different types of quantum channels.} Covariant: a channel with translational symmetry satisfying Eq.~\eqref{eq:covariance_cond}. Time-reversal: a single time-reversal channel $\tilde{\mathcal{N}}$ exists for classical and covariant cases, multiple channel  $\tilde{\mathcal{N}}^{\theta}$ coexist for incovariant cases.}\label{tab1}
\begin{tabular}{@{}cccc@{}}
\toprule
&Classical & Covariant& Incovariant\\
\midrule
\rule{0pt}{8pt}
$\omega$&real&real&complex\\
\rule{0pt}{10pt}
$P_{\to}\left( \omega \right)$&non-negative&real&complex\\
\rule{0pt}{10pt}
Coherence transfer&No&No&Yes\\
\rule{0pt}{15pt}
Time-reversal& $\begin{aligned} \tilde{\mathcal{N}}^{\theta} = \tilde{\mathcal{N}} \\[-4pt] ({\rm single}) \end{aligned}$ & $\begin{aligned} \tilde{\mathcal{N}}^{\theta} = \tilde{\mathcal{N}} \\[-4pt] ({\rm single}) \end{aligned}$ & $\begin{aligned} \tilde{\mathcal{N}}^{\theta} \neq \tilde{\mathcal{N}} \\[-4pt] ({\rm multiple}) \end{aligned}$ \\
\bottomrule
\end{tabular}
\end{table}

The covariance of a quantum channel is closely related to the multiplicity of its time-reversal process. A primitive form of the time-reversal quantum channel was introduced by Crooks \cite{PhysRevA.77.034101} as ${\cal \tilde N}(\hat\rho) = \sum_{x} \hat{K}^{\cal R}_{x} \hat{\rho}\hat{K}_{x}^{\cal R\dagger}$ with the time-reversal Kraus operator $\hat{K}^{\cal R}_{x} = \hat\gamma^{\frac{1}{2}}\hat{K}_{x}^{\dagger}\hat\gamma^{-\frac{1}{2}}$ when the forward process has the Kraus representation ${\cal N}(\hat\rho) = \sum_x \hat K_x \hat\rho \hat K_x^\dagger$ (see Methods). The time-reversal channel was later extended to include an additional degree of freedom ${\cal \tilde N}^\theta(\hat\rho) = \hat U_{\hat\gamma}^\dagger(\theta) {\cal \tilde {N}}( \hat U_{\hat\gamma}(\theta) \hat\rho \hat U_{\hat\gamma}^\dagger(\theta)) \hat U_{\hat\gamma}(\theta)$~\cite{junge_universal_2018}. We note that $\tilde{\mathcal{N}}$ is the special case with  $\theta = 0$, i.e., $\tilde{\mathcal{N}}=\tilde{\mathcal{N}}^{\theta = 0}$. This equation implies that multiple time-reversal channels can coexist when the forward channel ${\cal N}$ is incovariant under the group transformation $\hat{U}_{\hat\gamma}(\theta)$. In contrast, when ${\cal N}$ satisfies Eq.~\eqref{eq:covariance_cond}, all the time-reversal processes coincides, i.e., ${\cal \tilde N}^\theta = {\cal \tilde N}$ for every $\theta$.

While multiple time-reversal processes are possible in the quantum regime, a universal symmetry relation in Eq.~\eqref{Eq:Crooks_FT_Complex} can be found between the forward and each time-reversal process parametrized by $\theta$, regardless of the channel's covariance \cite{PhysRevX.9.031029}.  Moreover, the integral form of the QFT
\begin{equation}
\overline {e^{-\omega_R + 2 i \theta \omega_I } } = 1,
\end{equation}
for all $\theta \in (-\infty, \infty)$ plays a central role in the derivation of the second law of thermodynamics $\overline{\omega} = \sum_\omega \omega P_\rightarrow(\omega) \geq 0$ for the non-real quasi-probability distribution $P_\rightarrow(\omega)$~\cite{junge_universal_2018,PhysRevX.9.031029}.

\subsection*{Experimental demonstration}
\begin{figure*}[!t]
  \centering
  \includegraphics[width=0.99\linewidth]{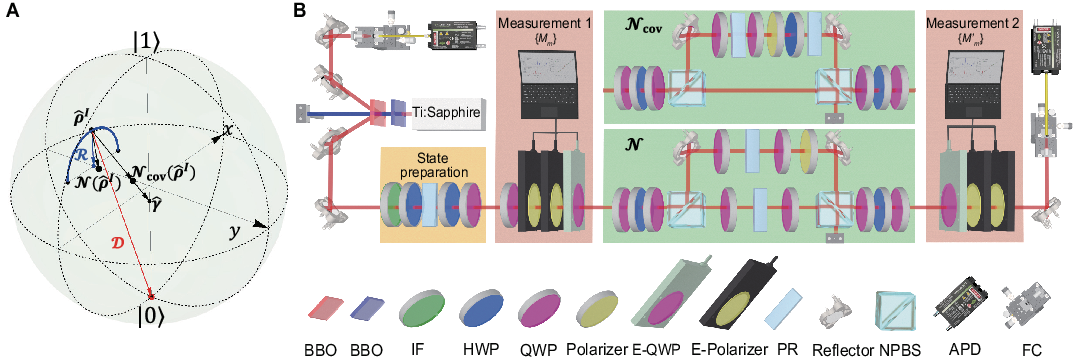}
	\caption{\textbf{Descriptions of quantum channels ${\cal N}$ and ${\cal N}_{\rm cov}$ in the Bloch sphere (A) and its experimental setup (B).} \textbf{(A)} Descriptions of quantum channels ${\cal N}$ and ${\cal N}_{\rm cov}$ in the Bloch sphere. The incovariant channel ${\cal N}$ is composed by mixing $\hat{\rho}$ and states $\mathcal{R}\left({\hat{\rho}} \right)$ and $\mathcal{D}\left( \hat{\rho} \right) $ after passing two processes ${\cal R}$ and ${\cal D}$ (see main text for details) ,while the covariant channel $\mathcal{N}_{\rm cov}$ is composed by mixing the input state $\hat{\rho}$ and the stationary state $\hat\gamma$. \textbf{(B)} Experimental setup. Pairs of photons are generated from spontaneous parametric down-conversion (SPDC) by pumping a beta-barium-borate (BBO) crystal. One photon is directly detected by an avalanche photodiode (APD) as the trigger, and the other is sent to the following setup including a state preparation module (orange shaded box), the two-point generalized measurement (two red-shaded boxes), and quantum channels (two green shaded boxes) and detected by another APD. The probability of the two-point generalized measurement outcomes $P(m,m')$ is obtained by the coincidence between two APDs. The abbreviations of the equipment are as follows: BBO, $\beta$-Barium borate crystal; IF, interference filter; HWP, half-wave plate; QWP, quarter-wave plate; E-QWP, electrically controlled quarter-wave plate; E-polarizer, electrically controlled polarizer; PR, phase retarder; NPBS, non-polarizing beam splitter; APD, avalanche photodiode; FC, fiber coupler.}
  \label{fig:schematic}
\end{figure*}

We experimentally demonstrate the generalized QFT applicable to both covariant and incovariant quantum channels
by reconstructing the quasi-probability distribution of the quantum entropy production using a quantum photonic setup. We encode a qubit system in the polarization DOF of a single photon by taking $\ket{0}$ and $\ket{1}$ as horizontally and vertically polarized states, respectively. In order to highlight the quantum signatures in entropy production, we design the following channel:
\begin{equation}\label{channel}
  \mathcal{N}(\hat\rho)=p\hat\rho+(1-p)[(1-s)\mathcal{R} (\hat\rho)+s\mathcal{D}(\hat\rho)],
\end{equation} 
where $\mathcal{R}$ is a mixture of two $\pm \pi/2$ rotations around the $y$-axis of the Bloch sphere and $\mathcal{D}$ maps any input states to $\ket{0}\bra{0}$ (see Fig.~\ref{fig:schematic}A).  The Kraus representation of $\mathcal{N}(\rho)$ is given in Methods. We note that the channel covers two types of decoherence: dephasing and amplitude damping, as $\mathcal{R} \left( \hat\rho \right)$ is a fully dephased state in $y$-basis. The stationary state of the channel is $\hat\gamma = (\frac{1 + s}{2}) \ket{0}\bra{0} + (\frac{1-s}{2}) \ket{1}\bra{1}$. Since the dephasing and amplitude damping processes are with respect to different axes, the channel ${\cal N}$ contains a non-trivial transition between off-diagonal elements, from $\ket{0}\bra{1}$ to $\ket{1}\bra{0}$. Thus, the channel does not meet the covariance condition.

To compare the covariant and incovariant quantum channels, we design the covariant counterpart of ${\cal N}$ satisfying Eq.~\eqref{eq:covariance_cond} with the form of  ${\cal{N}_{\rm cov}}\left( \hat{\rho} \right)  = p \hat{\rho}+\left( 1-p \right) \hat\gamma$, which shares the same stationary state with $\mathcal{N}$. The Kraus representation is given in Methods as well. This channel can be interpreted as statistically mixing the input state $\hat\rho$ and the stationary state $\hat\gamma$. In contrast to ${\cal N}$, the covariant channel ${\cal N}_{\rm cov}$ allows only the transitions from the initial off-diagonal element $\ket{0}\bra{1}$ and $|1\rangle \langle 0\vert$ to themselves, i.e., $\vert 0\rangle\langle 1\vert\to \vert 0\rangle \langle 1\vert$ and $\vert 1\rangle\langle 0\vert\to \vert 1\rangle \langle 0\vert$, while all other transitions such as transitions from off-diagonal elements to diagonal elements or transitions between off-diagonal elements are forbidden. Evidently, $\mathcal{N}_{\text{cov}}$ covers only the limited set of quantum channels such as thermalization while $\mathcal{N}$ is more general that it allows all possible transitions involving off-diagonal elements. For a quantum state only with diagonal elements in $\{ \ket{0}, \ket{1} \}$ basis, the two channels $\mathcal{N}$ and $\mathcal{N_{\rm cov}}$ lead to the same dynamics. More general forms of covariant channels can be found in the Supplementary Materials.

Another feature shared between ${\cal N}$ and ${\cal N}_{\rm cov}$ is that their time-reversal channels (for $\mathcal{N}$, one possible time-reversal channel) are the same as themselves, i.e., ${\cal \tilde N} =\tilde{\mathcal{N}}^{\theta= 0}= {\cal N}$ and ${\cal \tilde N}_{\rm cov}^\theta = {\cal \tilde N}_{\rm cov} = {\cal N}_{\rm cov}$. This time-reversal symmetry substantially reduces the experimental complexity without compromising the applicability of the theorem, since such symmetry is independent of the channel's covariance (see Methods). However, we note that there are other time-reversal channels ${\cal \tilde N}^\theta \neq {\cal N}$ as ${\cal N}$ is not covariant.

The experimental setup shown in Fig.~\ref{fig:schematic}B consists of four modules: the state preparation module, the quantum channel, and two measurement modules. In the preparation stage, a state $\hat\rho^I$ is prepared using two half-wave plates (HWPs), a quarter-wave plate (QWP), and phase retarders (PRs) which introduce time delays between two polarizations to control the mixture of the state. Two quantum channels $\mathcal{N}$ and  $\mathcal{N}_{\rm cov}$ are implemented using different combinations of two non-polarizing beam splitters (NPBSs), wave plates, and PRs. The first NPBS splits the input photon into two paths. For the implementation of $\mathcal{N}$, the transmitted path goes through the channel ${\cal R}^\varphi (\hat\rho) = \cos \varphi \hat\rho + (1- \cos \varphi) {\cal R} (\hat\rho)$ achieved by QWPs and PRs, which realizes the first two terms of $\mathcal{N}$, while for the covariant channel $\mathcal{N}_{\rm cov}$ the path passes directly with no operation. The reflected path is dissipated through the channel $\mathcal{D}$ and becomes $\ket{0}\bra{0}$ for ${\cal N}$ by passing through wave plates, PRs, and a  polarizer at a constant success probability independent of the input state, while for $\mathcal{N}_{\rm cov}$ extra wave plates and PRs are used to generate $\hat{\gamma}$ from $|0\rangle\langle 0\vert$. The reflected path then combines with the transmitted path at the second NPBS incoherently, achieving the desired channel $\mathcal{N}$ and $\mathcal{N}_{\rm cov}$.

In the experiment, the channel parameters are calibrated as $p=0.2864$ and $s=0.1316$, from which we can calculate the stationary state as $\hat\gamma= 0.5658\ket{0}\bra{0} + 0.4342 \ket{1}\bra{1}$. We first verify that $\hat{\gamma}$ is indeed the stationary state of $\mathcal{N}$ and  $\cal{N_{\rm cov}}$ by experimentally preparing  $\hat{\gamma}$ and charactering $\mathcal{N}\left( \hat{\gamma} \right) $ and $\cal{N}_{\rm cov}\left( \hat{\gamma} \right) $ via quantum state tomography. The results show high fidelity ($\approx 98.99\%$) between the three states. Through full quantum process tomography, the fidelity between the $\chi$-matrix of the reconstructed channel $\mathcal{N}$ and the theoretical one is $99.97\%$, and that for the channel $\mathcal{N}_{\rm cov}$ is $99.99\%$. These high fidelities surpass previous works \cite{HU20181551,PhysRevA.95.042310,PhysRevA.97.042112} and are crucial for the demonstration of the QFT. Such high fidelities are achieved by using auxiliary temporal and path DOFs of single photons, and deliberately designed coupling involving multiple DOFs to reduce the experimental complexity as well as imperfections. In addition, we use a rotating polarizer instead of the polarizing beam splitters (PBS) and HWP combination to improve the precision of the state preparation and the measurement (see Supplementary Materials for more details).

To investigate the QFT, we prepare the initial state $\hat\rho^I = \sum_\mu p^I_\mu \ket{\phi^I_\mu}\bra{\phi^I_\mu}$ with $p^I_0 = 4/5$, $p^I_1 = 1/5$, $\ket{\phi^I_0} =  \sin (\pi/6) \ket{0} - i \cos (\pi/6) \ket{1}$ and $\ket{\phi^I_1} = \cos (\pi/6) \ket{0} + i \sin (\pi/6) \ket{1}$, such that the quasi-probability distribution $P_\rightarrow(\omega)$  for the forward process only has real-values for both ${\cal N}$ and ${\cal N}_{\rm cov}$. We additionally perform the state tomography of the initial and final states, from which the entropy production is calculated according to Eq.~\eqref{eq:ent_prod}.

A major challenge in investigating QFTs is that a standard TPM protocol with the projection operator $\{\hat{\Pi}_i\}$ erases all the off-diagonal elements, thus preventing access to the transition amplitude $T^{\mu \rightarrow  \nu}_{ij \rightarrow kl}$.
We circumvent this problem by extending the TPM protocol to generalized measurements described by sets of operators $\{\hat{M}_m\}$ and $\{\hat{M}'_{m'} \}$ with outcomes $m$ and $m'$ before and after the state undergoes the quantum channel ${\cal N}$, respectively. The distribution of the measurement outcomes \cite{PhysRevX.9.031029,Wang2024} is given by $P(m,m') = {\rm Tr} [\hat{M}'_{m'} \mathcal{N}(\hat{M}_m \hat\rho \hat{M}_m^{\dag}) \hat{M}_{m'}^{\prime \dag}]$. We take appropriate measurement operators $\{ \hat{M}_m \} = \left\{ \frac{\hat{\Pi}_0 \hat{\Phi}^I_\mu}{\sqrt{2}}, \frac{\hat{\Pi}_1 \hat{\Phi}^I_\mu}{\sqrt{2}}, \frac{\hat{\Phi}^I_\mu}{2}, \frac{\hat{S}\hat{\Phi}^I_\mu}{2} \right\}_{\mu = 0,1}$ and $\{ \hat{M}'_{m'} \} = \left\{ \frac{\hat{\Phi}^F_{\nu}\hat{\Pi}_0 }{\sqrt{2}}, \frac{\hat{\Phi}^F_{\nu} \hat{\Pi}_1}{\sqrt{2}}, \frac{ \hat{\Phi}^F_{\nu}}{2}, \frac{\hat{\Phi}^F_{\nu} \hat{S}}{2} \right\}_{\nu = 0,1}$ using the combinations of $\hat{\Phi}^I_\mu$, $\hat{\Phi}^F_{\nu}$, $\hat{\Pi}_i$, and the phase gate $\hat{S} = \ket{0}\bra{0} + i \ket{1}\bra{1}$. The quasi-probability distribution can be obtained from a linear transform of $P(m,m')$ as $P_\rightarrow(\omega) = \sum_{m,m'} \alpha_{m m'}^\omega P(m,m')$ with some complex coefficients $\alpha_{m m'}^\omega$ (see Methods). In the experiment, the generalized measurements are realized by polarizers and QWPs before and after the channel followed by photodetection using an avalanche photodiode (APD). $P(m, m')$ is obtained by collecting the photon number statistics for different combinations of $m$ and $m'$. 

By utilizing the fact that ${\cal \tilde N} = {\cal N}$ and ${\cal \tilde N}_{\rm cov} = {\cal N}_{\rm cov}$, the time-reversal quasi-probability distribution $P^{\theta=0}_\leftarrow(\omega)=P_\leftarrow(\omega)$ for $\theta = 0$ is obtained by changing the input state to $\hat{\rho}^F_{\nu}$ and exchanging $\hat\Phi^I_\mu$ and $\hat\Phi^F_{\nu}$ in the measurement setting while keeping all the other configurations the same. 

We also note there exist some alternative approaches, including the weak measurement~\cite{PhysRevLett.108.070402} and interferometric protocols~\cite{PhysRevLett.113.140601, Gherardini_2018}, for estimating the quasi-probabilities $P_\rightarrow(\omega)$ and $P_\leftarrow^\theta(\omega)$ by regarding the transition amplitude $T^{\mu \rightarrow \nu}_{ij \rightarrow kl}$ as a variant of the Kirkwood-Dirac distribution (for more details, see recent review papers \cite{Lostaglio2023kirkwooddirac, arvidsson2024properties}.)

\begin{figure}[t]
	\centering
	\includegraphics[width=.48\textwidth]{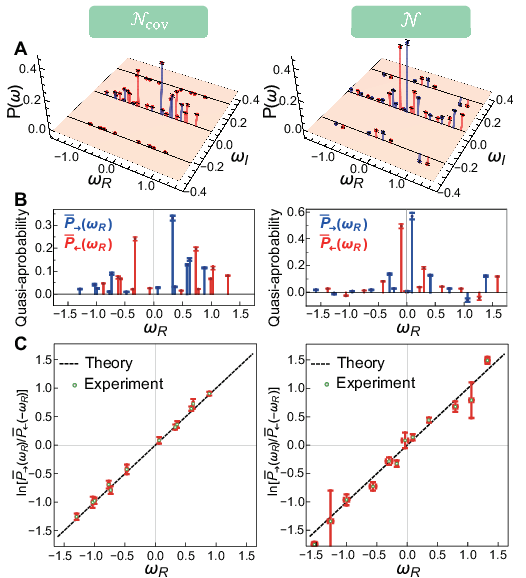}
	\caption{ \textbf{Reconstructed quasi-probability distributions.} \textbf{(A)} Theoretically predicted (bars) and experimentally reconstructed quasi-probability distributions (points) for the forward process $P_\rightarrow\left( \omega \right) $ (blue) and the time-reversal process $P_\leftarrow\left( \omega \right) $ (red) for ${\cal N}_{\rm cov}$ (left column) and  $\mathcal{N}$ (right column). \textbf{(B)} Quasi-probability distributions of the real part of the entropy production by averaging over all imaginary parts $\overline{P}_{\rightarrow}\left( \omega_R \right) = \sum_{\omega_I}P_{\rightarrow}\left( \omega_R + i\omega_I \right)$ (blue) and $\overline{P}_{\leftarrow}\left( \omega_R \right) = \sum_{\omega_I}P_{\leftarrow}\left( \omega_R + i\omega_I \right)$ (red) for ${\cal N}_{\rm cov}$ (left column) and ${\cal N}$ (right column). \textbf{(C)} The log-magnitude of the ratio between the forward and time-reversal quasi-probability distributions $\overline{P}_{\leftarrow}\left( \omega_{R}\right)/\overline{P}_{\rightarrow}\left( -\omega_{R} \right) $.}
 \label{fig:QFT_Re} 
\end{figure}

\subsection*{Verification of the QFT}

The reconstructed quasi-probability distributions $P_{\rightarrow}(\omega)$ and $P_{\leftarrow}(\omega)$ and the theoretical predictions for ${\cal N}_{\rm cov}$ and ${\cal N}$ are presented in Fig.~\ref{fig:QFT_Re}A. We observe that the deviations between the experimentally obtained quasi-probabilities and the theoretical ones $\sum_{\omega} \vert P_{\rightarrow \left( \leftarrow \right) }^{\rm \exp.}\left( \omega \right) - P_{\rightarrow\left( \leftarrow \right) }^{\rm theory}\left( \omega \right) \vert$ for both channels $\mathcal{N}$ and $\cal{N}_{\rm cov}$ are within $0.0734 \pm 0.0136$ (see Supplementary Materials for more details). The strong consistency between experimental results and the theoretical prediction demonstrates the high precision for the reconstruction of the quasi-probability distribution, which is indispensable for validating the QFT. The incovariant channel ${\cal N}$ can be clearly distinguished from the covariant channel ${\cal N}_{\rm cov}$ by its non-vanishing quasi-probability of the imaginary entropy production. This feature happens at $\omega_I = \pm \ln(\frac{1+s}{1-s}) \approx \pm 0.2647$, which corresponds to the incovariant transition between off-diagonal elements $\ket{0}\bra{1} \leftrightarrow \ket{1}\bra{0}$.

 \begin{figure*}[!htbp]
  \centering
  \includegraphics[width=0.99\textwidth]{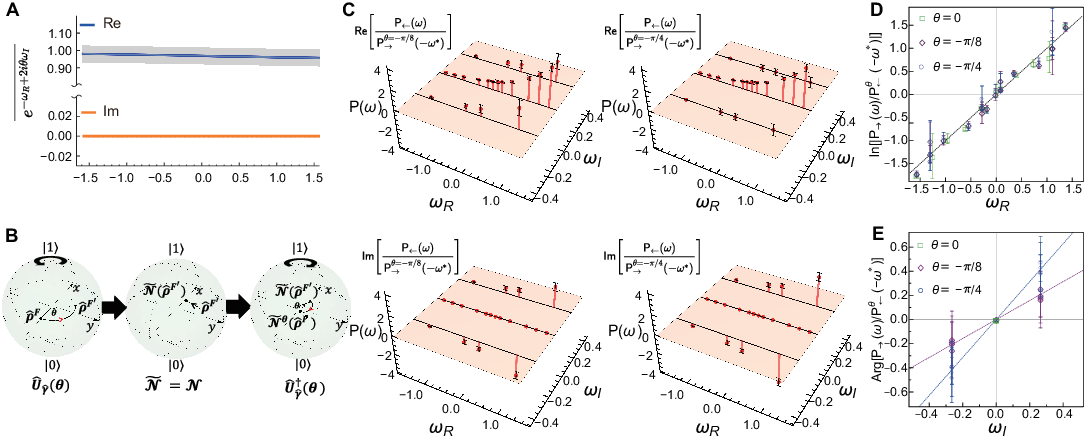}
	\caption{\textbf{Verification of the QFT.}
					\textbf{(A)}  Verification of integral fluctuation theorems $\overline{e^{-\omega_R + 2i\theta\omega_I }} = \sum_{\omega_R,\omega_I} P_\rightarrow(\omega) e^{-\omega_R + 2i\theta \omega_I}$ for $-\pi \leq \theta \leq \pi$ using experimentally reconstructed $P_{\rightarrow} (\omega)$ with $\omega = \omega_R + i \omega_I$. Blue and orange lines stand for the real part and the imaginary part of the integral, respectively, while the shaded gray represents the error bar. 
					\textbf{(B)} Description of rotated recovery map for the time-reversal process $\tilde{\mathcal{N}}^\theta$. The map comprises three components: 1) the group transformation $\hat{U}_{\hat{\gamma}}\left(  \theta\right)$ from the state $\hat{\rho}^{F}$ to the state $\hat{\rho}^{F^{\prime}} = \hat{U}_{\hat{\gamma}}\left( \theta \right)\hat{\rho} \hat{U}^{\dagger}_{\hat{\gamma}}\left( \theta \right) $, 2) the time-reversal channel $\tilde{\mathcal{N}}$, with $\tilde{\mathcal{N}} = \mathcal{N}$ in our experiment, and 3) the inverse group transformation $\hat{U}^{\dagger}_{\hat{\gamma}}\left( \theta \right)$ from the state $\tilde{\mathcal{N}}(\hat{\rho}^{F^{\prime}}) $ to $\tilde{\mathcal{N}}^{\theta}\left(\hat{ \rho}^{F} \right) = \hat{U}^{\dagger}_{\hat{\gamma}} \left( \theta \right)\tilde{\mathcal{N}}\left(\hat{U}_{\hat{\gamma}}\left( \theta \right)  \hat{\rho}^{F} \hat{U}^{\dagger}_{\hat{\gamma}}\left( \theta \right)  \right)\hat{U}_{\hat{\gamma}}\left( \theta \right)$.
					\textbf{(C)} Real parts (top) and imaginary parts (down) of the ratio between the quasi-probability distributions of entropy production of the forward and time-reversal processes. The red bars and points indicate the theoretical predictions and experimental values of the ratio, respectively. Experimental data are plotted with error bars.
					\textbf{(D, E)} Testing the QFT for both log-magnitude and phase of the quasi-probability ratio $P_\rightarrow(\omega)/P^\theta_\leftarrow(-\omega^*)$ with $\theta=0$ (green open square), $-\pi/8$ (purple open rhombus), and $-\pi/4$ (blue open circle). Black dashed line in \textbf{D} is the theoretical prediction for the real part  $\ln \vert P_\rightarrow(\omega)/P^\theta_\leftarrow(-\omega^*)\vert = \omega_R$, independent of $\omega_I$. Purple and blue dashed lines in \textbf{E} are the theoretical predictions for the imaginary part $\arg \vert P_\rightarrow(\omega)/P^\theta_\leftarrow(-\omega^*)\vert = -2\theta\omega_I$ for $\theta = -\pi/8$ and $\theta = -\pi/4$, respectively.
	Experimental data are plotted with error bars.}
  \label{fig:5}
\end{figure*}

\begin{table}[ht]
\centering
\caption{\textbf{Average entropy production $\overline{\omega} = \sum_{\omega}\omega P_{\to(\leftarrow)}^{\theta}\left( \omega \right)$ and integral fluctuation theorem $\overline{e^{\omega}} = \overline{e^{-\omega_R+2i\theta\omega_{I}}}$ when $\theta= 0$}\label{tab2}}
\begin{tabular}{@{}ccccc@{}}
\toprule
 & \multicolumn{2}{c}{Incovariant channel $\mathcal{N}$}& \multicolumn{2}{c}{Covariant channel $\mathcal{N}_{\text{cov}}$}\\
\midrule
 & Experiment& Theory& Experiment& Theory\\
\midrule
\rule{0pt}{10pt}
Average entropy production&$0.1447\pm0.0311$&$0.1182$&$0.2187\pm0.0142$&$0.2224$\\
\rule{0pt}{10pt}
Integral fluctuation theorem ($\theta = 0$)&$0.9699\pm0.0447$&$1$&$0.9887\pm0.0300$&$1$\\
\bottomrule
\end{tabular}
\end{table}

We first test the QFT for the time-reversal channel with $\theta=0$, in which case Eq.~\eqref{Eq:Crooks_FT_Complex} reduces to $P_\rightarrow (\omega) / P_\leftarrow (-\omega^*) = e^{\omega_R}$, independent of $\omega_I$. This fact allows us to focus only on the real part of entropy production by averaging over all imaginary parts, which yields a real-valued quasi-probability distribution $\overline {P}_{\rightarrow(\leftarrow)} (\omega_R) = \sum_{\omega_I} P_{\rightarrow(\leftarrow)} (\omega_R + i\omega_I)$ (see Fig.~\ref{fig:QFT_Re}B). We also observe negative values appearing in $\overline{P}_{\rightarrow}$ and $\overline{P}_{\leftarrow}$ for ${\cal N}$. Our experimental results clearly confirm that both channels obey a Crooks-like relation $\overline{P}_\rightarrow (\omega_R) / \overline P_\leftarrow (-\omega_R) = e^{\omega_R}$ as shown in Fig.~\ref{fig:QFT_Re}C.
We test the average entropy production $\overline{\omega}$ and the integral fluctuation theorem when $\theta =0$ for the incovariant and covariant channel (see Table \ref{tab2}). For a more general time-reversal channel with $\theta \neq 0$ of the incovariant channel ${\cal N}$, we first test the integral fluctuation theorem $\overline {e^{-\omega_R + 2 i \theta \omega_I } } = 1$ using the experimentally measured $P_{\rightarrow}\left( \omega \right) $, which holds for every $\theta$ (Fig.~\ref{fig:5}A). To test the Crooks-like relation with $\theta\neq 0$, we implement the unitary operation $\hat U_{\hat\gamma}(\theta) = \ket{0}\bra{0} + e^{i \ln (\frac{1+s}{1-s}) \theta} \ket{1}\bra{1}$ and $\hat{U}^{\dagger}_{\hat{\gamma}}\left( \theta \right)$ before and after $\tilde{\mathcal{N}}$ by realizing $z$-axis rotations with the QWP-HWP-QWP configurations (see Fig.~\ref{fig:5}B). We then take two specific values $\theta = -\pi/8$ and $\theta = -\pi/4$ for the time-reversal process ${\cal \tilde N}^\theta$ and experimentally measure the quasi-probability distribution $P_{\leftarrow}^{\theta}\left( \omega \right)$ (Fig.~\ref{fig:5}C) to demonstrate the fully quantum Crooks FT. To this end, we evaluate the ratio between the forward and time-reversal quasi-probability distributions for each $\omega = \omega_R + i \omega_I$, and take its log-magnitude and argument to test the relations $\ln \vert P_\rightarrow(\omega)/ P_\leftarrow^\theta(-\omega^*) \vert = \omega_R$ (Fig.~\ref{fig:5}D) and $\arg \left[ P_\rightarrow(\omega)/ P_\leftarrow^\theta(-\omega^*) \right] = -2i \theta \omega_I$ (Fig.~\ref{fig:5}E), which are equivalent to Eq.~\eqref{Eq:Crooks_FT_Complex}. The slopes obtained from the experimental data points $(\omega_R, \ln \lvert P_\rightarrow(\omega)/P_\leftarrow^\theta(-\omega^*) \rvert)$ are $1.04 \pm 0.08$ (for $\theta=0$), $1.03 \pm 0.06$ (for $\theta=-\pi/8$), and $0.98 \pm 0.07$ (for $\theta=-\pi/4$), which match the theory well. We also verify that the slopes obtained from the points $(\omega_I, {\rm arg}[P_\rightarrow(\omega)/P_\leftarrow^\theta(-\omega^*)])$ are $0.8 \pm 0.12$ (for $\theta=-\pi/8$) and $1.5 \pm 0.6$ (for $\theta=-\pi/4$), close to the values $(-2\theta)$ from the theory. These results confirm that while various choices can be made for the time-reversal processes of quantum channel, they all obey a quantum generalization of the Crooks FT.

\section*{Discussion}
In conclusion, we provide the experimental validation of generalized QFT in the presence of coherence using quantum photonic setups. The quasi-probability distributions of entropy production are reconstructed from the outcomes of the two-point generalized measurement protocols. The quantum generalization of the Crooks FT is successfully demonstrated for both covariant and incovariant quantum channels. Compared to the classical FTs, the non-classical characteristics of the QFT are highlighted by the imaginary part of entropy production originating from the multiple choices of the reverse channel for the incovariant channel and non-real quasi-probability distributions. 

Our experimental demonstration confirms the fundamental symmetry between a general quantum process and its time reverse with the notion of the imaginary part of entropy production. Formalisms and methodology introduced in this work will shed new light on investigating the role of coherence and noise during the quantum information processing applications~\cite{PhysRevLett.128.220502}, which will also lead to applications including quantum error correction \cite{doi:10.1063/1.1459754} and error suppression in continuous-time dynamics~\cite{Gertler2021,PhysRevLett.128.020403}.

\section*{Materials and Methods}
\subsection*{Entropy production for quantum channels}
Suppose a quantum system with time-independent Hamiltonian $\hat H$ is in thermal contact with a reservoir at temperature $T$. After a sufficiently long time, the system will arrive at the equilibrium state $\hat\gamma = \frac{e^{-\hat{H}/T} }{ {\rm Tr}[e^{-\hat{H} /T}]}$. When an initial quantum state $\hat\rho^I$ undergoes the thermalization process and evolves to the final state $\hat\rho^F$, the average entropy production can be expressed as $\overline \omega = \Delta S -Q/T$, where $\Delta S = S(\hat\rho^F) - S(\hat\rho^I)$ is the difference in the von Neumann entropy $S(\hat\rho) := -{\rm Tr} [\hat\rho \ln \hat\rho]$ and $Q = {\rm Tr} [(\hat\rho^F - \hat\rho^I) \hat{H}]$ corresponds to average heat transfer. The average entropy production can be rewritten as $\overline \omega = S(\hat\rho^I \| \hat\gamma) - S(\hat\rho^F \| \hat\gamma)$, in terms of the quantum relative entropy $S(\hat\rho \| \hat\gamma) := {\rm Tr} [\hat\rho (\ln\hat\rho - \ln \hat\gamma)]$ between $\hat\rho$ and the equilibrium state $\hat\gamma$. In other words, the non-negativity of average entropy production $\overline \omega \geq 0$ implies that the system is always getting close to its equilibrium state. 

Such an approach can be readily extended to a general non-unitary quantum channel ${\cal N}$ having a stationary state ${\cal N}(\hat\gamma) = \hat\gamma$. When an initial quantum state $\hat\rho^I$ evolves to ${\cal N}(\hat\rho^I) = \hat\rho^F$, average entropy production can be analogously defined as $\overline\omega = S(\hat\rho^I \| \hat\gamma) - S(\hat\rho^F \| \hat\gamma)= \Delta S - {\rm Tr} [(\hat\rho^F - \hat\rho^I) (-\ln \hat\gamma)] $, by replacing $\hat{H}/T$ with $(-\ln\hat\gamma)$, as known as a non-equilibrium potential \cite{PhysRevX.8.031037}. This formulation can be applied to a generic noisy quantum channel without having a well-defined temperature, where $\overline \omega \geq 0$ is regarded as a quantum generalization of the second law of thermodynamics \cite{Brandao}, guaranteed by the monotonicity of the quantum relative entropy.

\subsection*{Constructing quasi-probabilities of the quantum entropy production}

To explore FTs for a quantum process, we take a statistical point of view by regarding the initial and final states as the ensemble average of their eigenstates, $\hat\rho^I = \sum_\mu p^I_\mu \ket{\phi^I_\mu}\bra{\phi^I_\mu} = \sum_\mu p^I_\mu \hat\Phi_\mu^I$ and $\hat\rho^F = \sum_{\nu} p^F_{\nu} \ket{\phi^F_\nu}\bra{\phi^F_\nu} = \sum_\nu p_\nu^F \hat\Phi_\nu^F$. Here $p_{\mu \left( \nu \right) }^{I \left( F \right) }$ and $\left\vert \phi_{\mu \left( \nu \right) }^{I \left( F \right) }\right\rangle$ are the eigenvalues and eigenstates of the initial (final) quantum states, respectively. We assume that $\hat{\gamma} = \sum_{i}r_{i}\vert i\rangle \langle i \vert =\sum_{i}r_{i}\hat{\Pi}_{i}$ is the channel's stationary state such that ${\cal N}(\hat\gamma) = \hat\gamma$. The fluctuation of the quantum entropy production then can be explored by considering transitions between the eigenstates $\hat\Phi_\mu^I \rightarrow  \hat\Phi_\nu^F$, with probabilities $T^{\mu \rightarrow \nu} = {\rm Tr} [{\cal N} (\hat\Phi_\mu^I) \hat\Phi^F_\nu]$. The system's entropy change for each transition can be defined as $\left( \delta s  \right)^{\mu \rightarrow \nu} = \ln (p_\mu^I/ p^F_{\nu})$ so that averaging it over all possible transitions leads to the system's average entropy change, i.e., $\sum_{\mu, \nu} p^I_\mu\ T^{\mu \rightarrow \nu}\ \delta s^{\mu \rightarrow \nu} = \Delta S$.

A complex-valued transition amplitude can be designed to incorporate the transition between the off-diagonal elements as 
\begin{equation}
T^{\mu \rightarrow  \nu}_{ij \rightarrow kl} = {\rm Tr} \left[ {\cal N} \left(\hat{O}^I_{\mu i j} \right) \hat{O}^F_{\nu kl} \right],
\end{equation}
from $\hat{O}^I_{\mu i j} = \hat{\Pi}_i \hat{\Phi}^I_\mu \hat\Pi_j$ to $\hat{O}^F_{\nu kl}= \hat{\Pi}_k \hat{\Phi}^F_{\nu} \hat{\Pi}_l$. We highlight that the marginal distribution of $T^{\mu \rightarrow \nu}_{ij \rightarrow kl}$ reduces to proper transition probabilities of $T^{\mu \rightarrow \nu}$ and $T_{i \rightarrow k} = {\rm Tr} [{\cal N}(\hat\Pi_i) \hat\Pi_k]$.

For the time-reversal process ${\cal \tilde N}^\theta$, $P_\leftarrow^\theta(\omega)$ is obtained by exchanging $(p^I_\mu, \hat{O}^I_{\mu i j}) \leftrightarrow (p^F_\nu, \hat{O}^F_{\nu kl})$ and taking $\omega^{\nu \rightarrow \mu}_{kl \rightarrow ij} = -\omega^{\mu \rightarrow \nu}_{ij \rightarrow kl}$. 

One can check that the transition amplitude $T^{\mu \rightarrow  \nu}_{ij \rightarrow kl} = {\rm Tr} \left[ {\cal N} \left( \hat O^I_{\mu i j} \right) \hat O^F_{\nu kl} \right]$ contains the transition probabilities of both entropy change and information exchange. We note that the transition probability of the entropy change $T^{\mu \rightarrow \nu}$ can be obtained by adding up all the indices but leaving $\mu$ and $\nu$ as

\begin{equation}
\begin{aligned}
T^{\mu \rightarrow \nu} &= \sum_{i,j,k,l} T^{\mu \rightarrow  \nu}_{ij \rightarrow kl} \\
&= \sum_{i,j,k,l}  {\rm Tr} \left[ {\cal N} \left( \hat{O}^I_{\mu i j} \right) \hat{O}^F_{\nu kl} \right] \\
&= \sum_{i,j,k,l}  {\rm Tr} \left[ {\cal N} \left(\hat\Pi_i\hat{\Phi}_\mu^I \hat\Pi_j \right) \hat\Pi_k \hat{\Phi}_\nu^I \hat\Pi_l \right] \\
&=  {\rm Tr} \left[ {\cal N} (\hat{\Phi}_\mu^I) \hat{\Phi}_\nu^I  \right],
\end{aligned}
\end{equation}
from the completeness relation $\sum_i \hat{\Pi}_i = \mathbb{1}$, where $\mathbb{1}$ is the identity operator. The transition probability of the information exchange for $i\rightarrow k$ can be obtained in a similar manner as

\begin{equation}
\begin{aligned}
T_{i \rightarrow k} &= \sum_{\mu, \nu, j, l} T^{\mu \rightarrow  \nu}_{ij \rightarrow kl} \\
&= \sum_{\mu, \nu, j, l}  {\rm Tr} \left[ {\cal N} \left( \hat{O}^I_{\mu i j} \right) \hat{O}^F_{\nu kl} \right] \\
&= \sum_{\mu, \nu, j, l} {\rm Tr} \left[ {\cal N} \left(\hat\Pi_i\hat{\Phi}_\mu^I \hat\Pi_j \right) \hat\Pi_k \hat{\Phi}_\nu^I \hat\Pi_l \right] \\
&= {\rm Tr} [{\cal N}(\hat\Pi_i) \hat\Pi_k],
\end{aligned}
\end{equation}
from $\sum_\mu \hat{\Phi}_\mu^I = \mathbb{1}$, $\sum_\nu \hat{\Phi}_\nu^F = \mathbb{1}$ and $\Pi_i \Pi_j = \delta_{i j} \Pi_i$.

We show that the average entropy production becomes $\overline\omega = \sum_\omega \omega P_\rightarrow(\omega) = S(\hat\rho^I \| \hat\gamma) - S(\hat\rho^F \| \hat\gamma)$. From Eq.~\eqref{eq:entprod_dist}, the average entropy production can be written as $\overline\omega = \sum_\omega \omega P_\rightarrow(\omega) = \sum_{\mu, \nu,i,j,k,l} p^{I}_{\mu}T^{\mu\rightarrow \nu}_{ij\rightarrow kl} \omega^{\mu \rightarrow \nu}_{ij \rightarrow kl}$, in terms of the complex-valued transition amplitude $T^{\mu \rightarrow  \nu}_{ij \rightarrow kl} $ and the stochastic entropy production $\omega^{\mu\rightarrow \nu}_{ij\rightarrow kl}$ in Eq.~\eqref{eq:ent_prod}. By defining $\langle \xi \rangle = \sum_{\mu, \nu,i,j,k,l} p^{I}_{\mu}T^{\mu\rightarrow \nu}_{ij\rightarrow kl} \xi$ as the average of stochastic variable $\xi$ over all possible transitions, the real and imaginary parts of the average entropy production then can be expressed as 
\begin{equation}
\overline\omega_R 
= \left\langle \ln\left( \frac{p^{I}_{\mu}\sqrt{r_{k}r_{l}} }{p_{\nu}^{F}\sqrt{r_{i}r_{j}} } \right) \right\rangle
\end{equation}
and
\begin{equation}
\overline\omega_I
= \left\langle  \ln\left( \frac{ \sqrt{r_{j}r_{l} }}{ \sqrt{r_{i}r_{k}} }  \right) \right\rangle,
\end{equation}
respectively. From the completeness relation (see Supplementary Materials for more details), we note that
\begin{equation}
\begin{aligned}
\langle \ln p^I_\mu \rangle & = {\rm Tr}[\hat\rho^I \ln \hat\rho^I],\\
\langle \ln p^F_\nu \rangle &= {\rm Tr}[\hat\rho^F \ln \hat\rho^F],\\
\langle \ln r_i \rangle &= {\rm Tr}[\hat\rho^I \ln \hat\gamma] = \langle \ln r_j \rangle.
\end{aligned}
\end{equation}
This leads to $\overline \omega_R = {\rm Tr} [\hat{\rho}^I \ln \hat{\rho}^I]  - {\rm Tr} [\hat{\rho}^F \ln \hat{\rho}^F] + {\rm Tr}[\hat{\rho}^F\ln \hat{\gamma}] - {\rm Tr}[\hat{\rho}^I \ln \hat{\gamma}] = S(\hat{\rho}^I \| \hat{\gamma}) - S(\hat{\rho}^F \| \hat{\gamma})$ and $\overline \omega_I = 0$, which completes the proof.

\subsection*{Kraus representations of the incovariant and covariant quantum channels}
The Kraus representation of the incovariant process $\mathcal{N}$ in Eq.~\eqref{channel} is given by $\mathcal{N}\left( \hat{\rho} \right) =\sum_{x=0}^4 \hat{K}_{x} \hat{\rho}\hat{K}_{x}^{\dagger}$ with the following Kraus operators:
\begin{equation}
\begin{aligned}
				\widehat{K}_0&=\sqrt{p}\begin{pmatrix}1&0\\0&1\end{pmatrix}, \\
				\widehat{K}_1&=\frac{\sqrt{1-p}\sqrt{1-s}}{2}\begin{pmatrix}1&1\\-1&1\end{pmatrix}, \\
				\widehat{K}_{2}&=\frac{\sqrt{1-p}\sqrt{1-s}}{2}\begin{pmatrix}1&-1\\1&1\end{pmatrix}, \\
				\widehat{K}_3&=\sqrt{1-p}\sqrt{s}\begin{pmatrix}0&0\\1&0\end{pmatrix}, \\
				\widehat{K}_4&=\sqrt{1-p}\sqrt{s}\begin{pmatrix}0&0\\0&1\end{pmatrix}. 
\end{aligned}
\end{equation}

Similarly, the Kraus representation of the covariant process $\mathcal{N}_{\text{cov}} = p \hat\rho + (1-p) \hat\gamma$ is given by $\mathcal{N}_{\text{cov}}\left( \hat{\rho} \right) =\sum_{x=0}^4 \hat{K}_{x\text{cov}}\hat{\rho}\hat{K}^{\dagger}_{x\text{cov}} $ with the following Kraus operators:
\begin{equation}
\begin{aligned}
				\widehat{K}_{0\text{cov}}&=\sqrt{p}\begin{pmatrix}1&0\\0&1\end{pmatrix}, \\
				\widehat{K}_{\text{1cov}}&=\sqrt{\frac{\left( 1-p \right) \left( 1-s \right) }{2}}\begin{pmatrix}1&0\\0&0\end{pmatrix}, \\
				\widehat{K}_{2\text{cov}}&=\sqrt{\frac{\left( 1-p \right) \left( 1+s \right) }{2}}\begin{pmatrix}0&0\\1&0\end{pmatrix}, \\
				\widehat{K}_{3\text{cov}}&=\sqrt{\frac{\left( 1-p \right) \left( 1-s \right) }{2}}\begin{pmatrix}0&1\\0&0\end{pmatrix}, \\
				\widehat{K}_{\text{4cov}}&=\sqrt{\frac{\left( 1-p \right) \left( 1+s \right) }{2}}\begin{pmatrix}0&0\\0&1\end{pmatrix}. 
\end{aligned}
\end{equation}

More details on the derivation of the Kraus representation of the incovariant process and the covariant process can be found in the Supplementary Materials

\subsection*{The time-reversal quantum process}
We confirm the validity of the time-reversal quantum channel defined as
\begin{equation}
\begin{aligned}
{\cal \tilde N}(\hat\rho) &= \sum_x (\hat\gamma^{\frac{1}{2}} \hat K^\dagger_x \hat\gamma^{-\frac{1}{2}}) \hat\rho (\hat\gamma^{-\frac{1}{2}} \hat K_x \hat\gamma^{\frac{1}{2}})= \sum_x \hat K^{\cal R}_x \hat\rho \hat K^{\cal R \dagger}_x,
\end{aligned}
\end{equation}
where $\hat K^{\cal R}_x = \hat\gamma^{\frac{1}{2}} \hat K^\dagger_x \hat\gamma^{-\frac{1}{2}}$ and $\hat K^{\cal R \dagger}_x = \hat\gamma^{-\frac{1}{2}} \hat K_x \hat\gamma^{\frac{1}{2}}$. We first note that $\tilde {\cal N}$ is a valid quantum channel described by the Kraus operator $\hat K^{\cal R}_x$, satisfying $\sum_x \hat K_x^{\cal R\dagger} \hat K^{\cal R}_x = \sum_x (\hat\gamma^{-\frac{1}{2}} \hat K_x \hat\gamma^{\frac{1}{2}})  (\hat\gamma^{\frac{1}{2}} \hat K^\dagger_x \hat\gamma^{-\frac{1}{2}}) = \sum_x (\hat\gamma^{-\frac{1}{2}} \hat K_x \hat\gamma \hat K^\dagger_x \hat\gamma^{-\frac{1}{2}}) = \hat\gamma^{-\frac{1}{2}} \sum_x (\hat K_x \hat\gamma \hat K^\dagger_x) \hat\gamma^{-\frac{1}{2}}  = \hat\gamma^{-\frac{1}{2}} \hat\gamma \hat\gamma^{-\frac{1}{2}} = \mathbb{1}$ from the definition of the stationary state ${\cal N} (\hat\gamma) = \sum_x \hat K_x \hat\gamma \hat K_x^\dagger = \hat\gamma$. We can easily check that ${\cal \tilde N} (\hat\gamma) = \hat \gamma$ so that $\hat \gamma$ is also a fixed point of the time-reversal channel. 

Moreover, sequential applications of the Kraus operators lead to $p( x_1, x_2, \cdots, x_n \vert \hat\gamma) = {\rm Tr} [\hat K_{x_n} \cdots \hat K_{x_2} \hat K_{x_1} \hat\gamma \hat K_{x_1}^\dagger \hat K_{x_2}^\dagger \cdots \hat K_{x_n}^\dagger]$ for the forward process. By applying the Kraus operators for the time-reversal process, we note that ${\rm Tr} [\hat K^{\cal R}_{x_1}  \hat K^{\cal R}_{x_2} \cdots \hat K^{\cal R}_{x_n} \hat\gamma \hat K^{\cal R \dagger}_{x_n} \cdots \hat K^{\cal R \dagger}_{x_2} \hat K^{\cal R \dagger}_{x_1}] = \\\tilde p( x_n, \cdots, x_2, x_1 \vert \hat\gamma) = p( x_1, x_2, \cdots, x_n \vert \hat\gamma)$, which is given in the reverse order compared to the forward probabilities. Such a property can be regarded as a quantum mechanical generalization of classical Markov chain time-reversal \cite{PhysRevA.77.034101}.

By introducing an additional rotation DOF $\theta$ with $\hat{U}_{\hat\gamma}(\theta) = e^{-i\theta \ln \hat\gamma} = \hat\gamma^{-i\theta}$, a general family of the time-reversal channels is defined as
\begin{equation}
\begin{aligned}
{\cal \tilde N}^\theta(\hat\rho) 
&= \hat U_{\hat\gamma}^\dagger(\theta) {\cal \tilde {N}}( \hat U_{\hat\gamma}(\theta) \hat\rho \hat U_{\hat\gamma}^\dagger(\theta)) \hat U_{\hat\gamma}(\theta)\\
&= \sum_x (\hat\gamma^{\frac{1}{2}+i\theta} \hat K^\dagger_x \hat\gamma^{-\frac{1}{2}-i\theta}) \hat\rho (\hat\gamma^{-\frac{1}{2}+i\theta} \hat K_x \hat\gamma^{\frac{1}{2}-i\theta}) \\
&= \sum_x \hat K_x^{\cal R\theta} \hat\rho \hat K_x^{\cal R\theta \dagger},
\end{aligned}
\end{equation}
by defining $\hat K_x^{\cal R\theta} = \hat\gamma^{\frac{1}{2}+i\theta} \hat K^\dagger_x \hat\gamma^{-\frac{1}{2}-i\theta}$. This family incorporates the standard time-reversal channel ${\cal \tilde N}$ as the special case of $\theta=0$. Similarly to ${\cal \tilde N}$, ${\cal \tilde N}^\theta$ satisfies all the properties described above.

\subsection*{Independence of time-reversal symmetry on channel's covariance}
We discuss that imposing time-reversal symmetry, ${\cal \tilde{N}} = {\cal \tilde{N}}^{\theta=0} = {\cal N}$, in such a way that one of the reverse channels coincides with the forward channel does not trivialize the QFT, as it does not reduce to the covariance condition. To show this explicitly, we rewrite the condition of the time-reversal symmetry in terms of the eigenbasis of the stationary state as 
\begin{equation}
\begin{aligned}
\bra{k} {\cal N}(\ket{i}\bra{j}) \ket{l} &= \bra{k} {\cal \tilde{N}}(\ket{i}\bra{j}) \ket{l} \\
&= \bra{k} \sum_x (\hat\gamma^{\frac{1}{2}} \hat K^\dagger_x \hat\gamma^{-\frac{1}{2}}) \ket{i}\bra{j}(\hat\gamma^{-\frac{1}{2}} \hat K_x \hat\gamma^{\frac{1}{2}}) \ket{l} \\
&= \sqrt{\frac{r_k r_l}{r_i r_j}} \bra{j} {\cal N}(\ket{l}\bra{k}) \ket{i}.
\end{aligned}
\end{equation}
This implies that the transition amplitudes $T_{ij \rightarrow kl}=\bra{k} {\cal N}(\ket{i}\bra{j}) \ket{l}$ and $T_{lk \rightarrow ji}=\bra{j} {\cal N}(\ket{l}\bra{k}) \ket{i}$ satisfy the ratio $\sqrt{r_i r_j} T_{ij \rightarrow kl} = \sqrt{r_k r_l} T_{lk \rightarrow ji}$. However, it is worth noting that such a condition does not prohibit transitions between off-diagonal elements.

On the other hand, the covariance condition in Eq.~\eqref{eq:covariance_cond} implies
\begin{equation}
e^{-i\theta (\ln r_i - \ln r_j)} \bra{k} {\cal N}(\ket{i}\bra{j}) \ket{l} = e^{-i\theta (\ln r_k - \ln r_l)} \bra{k} {\cal N}(\ket{i}\bra{j}) \ket{l},
\end{equation}
for any $\theta \in \mathbb{R}$, so that the transition $T_{ij \rightarrow kl}$ is possible only if $\ln r_i - \ln r_j = \ln r_k - \ln r_l$. This condition imposes a strong restriction, forbidding certain transitions between off-diagonal elements. More explicitly, for a two-level system with $r_0 \neq r_1$, the allowed transitions between off-diagonal elements by a covariant channel are $T_{01 \rightarrow 01}$ and $T_{10 \rightarrow 10}$, i.e., transitions to themselves, while transitions to different off-diagonal elements, $T_{01 \rightarrow 10}$ and $T_{10 \rightarrow 01}$ are forbidden.

Therefore, time-reversal symmetry and covariance are independent conditions from each other, and all essential features of generalized fluctuation theorems for incovariant channels, such as imaginary entropy production, can still be observed even when considering a channel with time-reversal symmetry.

\subsection*{Reconstructing the quasi-probability distribution from generalized measurements}
We show that the quasi-probability distribution of the entropy production can be reconstructed from the generalized measurement operators $\{ \hat{M}_m \} = \left\{ \frac{ \hat{\Pi}_{0} \hat{\Phi}^I_\mu}{\sqrt{2}}, \frac{ \hat{\Pi}_1 \hat{\Phi}^I_\mu}{\sqrt{2}}, \frac{ \hat{\Phi}^I_\mu}{2}, \frac{ \hat{S} \hat{\Phi}^I_\mu}{2} \right\}_{\mu = 0,1}$ and $\{ \hat{M}'_{m'} \} = \left\{ \frac{\hat{\Phi}^F_{\nu} \hat{\Pi}_0 }{\sqrt{2}}, \frac{ \hat{\Phi}^F_{\nu} \hat{\Pi}_1}{\sqrt{2}}, \frac{ \hat{\Phi}^F_{\nu}}{2}, \frac{ \hat{\Phi}^F_{\nu} \hat{S}}{2} \right\}_{\nu = 0,1}$, performed before and after that the state undergoes the quantum channel. For simplicity, let us express the measurement operators as
\begin{equation}
\begin{aligned}
\hat{M}_m &= \hat{M}_{(\mu, r)} = \hat{L}_r \hat{\Phi}^I_\mu ,\\
\hat{ M}'_{m'} &= \hat{M}'_{(\nu, s)} = \hat{ \Phi}^F_\nu \hat{L}_s,
\end{aligned}
\end{equation}
by defining $\hat{ L}_1 = \frac{ \hat{\Pi}_0}{\sqrt{2}}$, $L_2 = \frac{\hat{\Pi}_1}{\sqrt{2}}$, $\hat{L}_3 = \frac{\mathbb{1}}{2}$, and $\hat{L}_4 = \frac{\hat{S}}{2}$. Denominators in $\hat{L}_r$ are chosen to ensure the completeness condition $\sum_m \hat{M}^\dagger_m \hat{M}_m = \mathbb{1}$ and $\sum_{m'} \hat{M}'^\dagger_{m'} \hat{M}'_{m'} = \mathbb{1}$. As $\hat{\Phi}^I_\mu$ and $\hat{ \Phi}^F_\nu$ are projectors onto two different eigenstates with $\mu=\{0,1\}$ and $\nu=\{0,1\}$, respectively, each measurement has $2 \times 4 = 8$ outcomes. As a combination, the two-point measurement has $8\times 8 = 64$ measurement outcomes. To reconstruct the quasi-probability distribution, we first note that 
\begin{equation}
\hat{ M}_{(\mu,r)} \hat\rho^I \hat{M}_{(\mu,r)}^\dagger = p^I_\mu \hat{L}_r \hat{\Phi}^I_\mu \hat{L}_r^\dagger.
\end{equation}

We then used the fact that any operator $\hat{A}$ described by a $2\times2$ matrix satisfies the following relation:
\begin{equation}
\label{eq:c_r_table}
\begin{aligned}
\hat{\Pi}_0 \hat{A} \hat{\Pi}_0 &= 2  \hat{L}_1 \hat{A} \hat{L}_1^\dagger,\\
\hat{\Pi}_0 \hat{A} \hat{\Pi}_1 &= -(1+i) \hat{L}_1 \hat{A} \hat{L}_1^\dagger -(1+i) \hat{L}_2 \hat{A} \hat{L}_2^\dagger + 2 \hat{L}_3 \hat{A} \hat{L}_3^\dagger + 2i \hat{L}_4 \hat{A} \hat{L}_4^\dagger ,\\
\hat{\Pi}_1 \hat{A} \hat{\Pi}_0 &= -(1-i) \hat{L}_1 \hat{A} \hat{L}_1^\dagger -(1-i) \hat{L}_2 \hat{A} \hat{L}_2^\dagger + 2 \hat{L}_3 \hat{A} \hat{L}_3^\dagger - 2i \hat{L}_4 \hat{A} \hat{L}_4^\dagger, \\
\hat{\Pi}_1 \hat{A} \hat{\Pi}_1 &=  2 \hat{L}_2 \hat{A} \hat{L}_2^\dagger,
\end{aligned}
\end{equation}
which can be summarized as $\hat{\Pi}_i \hat{A} \hat{\Pi}_j = \sum_{r} c^{ij}_r  \hat{L}_r \hat{A} \hat{L}_r^\dagger$  with the complex coefficients $c_r^{ij}$. Therefore, one can express the operator at the initial point
\begin{equation} \label{eq:Oij}
\begin{aligned}
p^I_\mu \hat{O}^I_{\mu ij} &= p^I_\mu \hat{\Pi}_i \hat{\Phi}^I_\mu \hat{\Pi}_j = \sum_{r} c_r^{ij} p^I_\mu \hat{L}_r \hat{\Phi}^I_\mu \hat{L}_r^\dagger = \sum_r c^{ij}_r \hat{M}_{(\mu,r)} \hat\rho^I \hat{M}^\dagger_{(\mu,r)}
\end{aligned}
\end{equation}
as a linear combination of the effects of the first measurement. For the second set of measurement operators, we observe
\begin{equation}
\hat{M}'^\dagger_{(\nu,s)} \hat{M}'_{(\nu,s)} = \hat{L}_s^\dagger \hat{\Phi}^F_\nu \hat{L}_s.
\end{equation}
Similarly, we can find the following relation for any operator $\hat{A}$:
\begin{equation}
\label{eq:c_s_table}
\begin{aligned}
\hat{\Pi}_0 \hat{A} \hat{\Pi}_0 & = 2 \hat{L}_1^\dagger \hat{A} \hat{L}_1,\\
\hat{\Pi}_0 \hat{A} \hat{\Pi}_1 & = -(1-i) \hat{L}_1^\dagger \hat{A} \hat{L}_1 -(1-i) \hat{L}_2^\dagger \hat{A} \hat{L}_2 + 2 \hat{L}_3^\dagger \hat{A} \hat{L}_3 - 2i \hat{L}_4^\dagger \hat{A} \hat{L}_4, \\
\hat{\Pi}_1 \hat{A} \hat{\Pi}_0 & = -(1+i) \hat{L}_1^\dagger \hat{A} \hat{L}_1-(1+i) \hat{L}_2^\dagger \hat{A} \hat{L}_2 + 2 \hat{L}_3^\dagger \hat{A} \hat{L}_3 + 2i \hat{L}_4^\dagger \hat{A} \hat{L}_4,\\
\hat{\Pi}_1 \hat{A} \hat{\Pi}_1 & = 2 \hat{L}_2^\dagger \hat{A} \hat{L}_2, 
\end{aligned}
\end{equation}
or equivalently, $\hat{\Pi}_i \hat{A} \hat{\Pi}_j = \sum_r d^{ij}_r \hat{L}_r^\dagger \hat{A} \hat{L}_r = \sum_r c^{ji}_r \hat{L}_r^\dagger \hat{A} \hat{L}_r$ by comparing Eq.~\eqref{eq:c_r_table} and \eqref{eq:c_s_table}. We can express the operator at the final point
\begin{equation} \label{eq:Okl}
\begin{aligned}
\hat{O}^F_{\nu k l} &= \hat{\Pi}_k \hat{\Phi}_\nu^F \hat{\Pi}_l  = \sum_s c^{lk}_s \hat{L}_s^\dagger \hat{\Phi}^F_\nu \hat{L}_s = \sum_s c^{lk}_s  \hat{M}'^\dagger_{(\nu,s)} \hat{M}'_{(\nu,s)}.
\end{aligned}
\end{equation}
By combining Eq.~\eqref{eq:Oij} and \eqref{eq:Okl}, we obtain
\begin{equation}
\begin{aligned}
p^I_\mu T^{\mu \rightarrow  \nu}_{ij \rightarrow kl} 
&= {\rm Tr} \left[ {\cal N} (p^I \hat{O}^I_{\mu i j} ) \hat{O}^F_{\nu k l} \right] \\
&= {\rm Tr} \big[  {\cal N} \big ( \sum_r c^{ij}_r \hat{M}_{(\mu,r)} \hat\rho^I \hat{M}^\dagger_{(\mu,r)} \big ) \big( \sum_s c^{lk}_s \hat{M}'^\dagger_{(\nu,s)} \hat{M}'_{(\nu,s)} \big) \big]\\
&= \sum_{r,s} c^{ij}_r c^{lk}_s {\rm Tr} \left[ \hat{M}'_{(\nu,s)}\right. {\cal N} \left.\left( \hat{M}_{(\mu,r)}\hat\rho^I \hat{M}^\dagger_{(\mu,r)}\right) \hat{M}'^\dagger_{(\nu,s)}\right]\\
&= \sum_{r,s} c^{ij}_r c^{lk}_s P (m,m'),
\end{aligned}
\end{equation}
where $m = (\mu, r)$ and $m' = (\nu, s)$. This result directly leads to 
\begin{equation}
\begin{aligned}
P_\rightarrow(\omega) &= \sum_{\mu,\nu,i,j,k,l} \delta(\omega - \omega^{\mu \rightarrow \nu}_{ij \rightarrow kl} ) p^I_\mu T^{\mu \rightarrow  \nu}_{ij \rightarrow kl}  \\
											&=\sum_{\mu,\nu,i,j,k,l} \delta(\omega - \omega^{\mu \rightarrow \nu}_{ij \rightarrow kl} ) \sum_{r,s} c^{ij}_r c^{lk}_s P (m,m')\\
&=\sum_{\mu, r} \sum_{\nu,s} \sum_{i,j,k,l} \delta(\omega - \omega^{\mu \rightarrow \nu}_{ij \rightarrow kl} ) c^{ij}_r c^{lk}_s P (m,m')\\
&=\sum_{m,m'}  \sum_{i,j,k,l} \delta(\omega - \omega^{\mu \rightarrow \nu}_{ij \rightarrow kl} ) c^{ij}_r c^{lk}_s P (m,m')\\
&=\sum_{m,m'}  \alpha^\omega_{mm'} P (m,m'),
\end{aligned}
\end{equation}
by defining $\alpha^\omega_{mm'} := \sum_{i,j,k,l} \delta(\omega - \omega^{\mu \rightarrow \nu}_{ij \rightarrow kl} )c^{ij}_r c^{lk}_s$ and noting that $\sum_m (\cdot)= \sum_{\mu,r}(\cdot)$ and $\sum_{m'} (\cdot) = \sum_{\nu,s}(\cdot)$.

%

\section*{Supplementary Materials}
The Supplementary Materials include:\\
Supplementary Text Section S1 to S3\\
Figs. S1 to S12\\
Table S1

\noindent \textbf{Acknowledgements:} 

%
\noindent \textbf{Funding:} This work was supported by the National Natural Science Foundation of China (Grants No. 12347104, No. U24A2017, No. 12461160276), National Key Research and Development Program of China (Grants No. 2023YFC2205802), Natural Science Foundation of Jiangsu Province (Grants No. BK20243060, BK20233001), in part by State Key Laboratory of Advanced Optical Communication Systems and Networks, China. H.K. and M.S.K. are supported by the National Research Foundation of Korea (NRF) Grant funded by the Korea government (MIST) (Grant No. RS-2024-00413957). H.K. is supported by the National Research Foundation of Korea (NRF) grant funded by the Korea government (MSIT)(No. RS-2024-00438415) and the KIAS individual grant No. CG085302 at the Korea Institute for Advanced Study. M.S.K. is supported by the UK EPSRC through EP/W032643/1, EPY004752/1 and EP/Z55318X/1 and acknowledges the KIST Open Research program and KIAS visiting professorship.\\
\noindent \textbf{Author Contributions:} With the help of H.K., Y.X.Z., J.X., H.L., M.S.K., and L.J.Z. conceived and designed the experiment, H.L. J.X. and Y.X.Z. performed the experiment, H.L. and J.X. analysed the data, H.L., J.X. and H.K. performed simulations of the experiment and co-wrote the paper. L.J.Z. and M.S.K. supervised the work.  \\
\noindent \textbf{Competing Interests:} The authors declare that they have no competing interests.\\
\noindent \textbf{Data and materials availability:} All data needed to evaluate the conclusions in the paper are present in the paper and/or the Supplementary Materials.


\end{document}